\documentclass[prb,preprint,groupedaddress]{revtex4}
\usepackage{graphics}
\def\trc{TRC }
\def\e{\mathcal{E}}
\begin{document}
\author{X.-N. Li}
\author{J. Machta}
\email[]{machta@physics.umass.edu}
\affiliation{Department of Physics and Astronomy,
University of Massachusetts, Amherst,
MA 01003-3720}
\begin{abstract}

The dynamic critical behavior of the two-replica cluster algorithm is studied. Several
versions of the algorithm are applied to the two-dimensional, square lattice Ising
model with a staggered field. The dynamic exponent for the full algorithm is 
found to be less than 0.4. It is found that odd translations of one replica with respect to
the other together with global flips are essential for obtaining a small value of the dynamic
exponent.

\end{abstract}
\pacs{}

\title{Critical dynamics of two-replica cluster algorithms}

\maketitle

\section{Introduction}

The Swendsen-Wang (SW) algorithm and related cluster
methods~\cite{SwWa,Wolff,KaDo,ChMa97a,ChMa98a,WaSw90,NeBa99}  have greatly improved the efficiency
of simulating the critical region of a variety of spin models.  The original
SW algorithm can be modified to work for spin systems with internal
symmetry breaking fields~\cite{DoSeTa}.  Spin models of this kind include the Ising antiferromagnet in a uniform
field, the random field Ising model and lattice gas models of adsorption in porous media~\cite{DuMaSaAu}. The
modification proposed in Ref. \onlinecite{DoSeTa} is to assign Boltzmann weights depending on the
net field acting on the cluster to decide whether the cluster should be flipped. Unfortunately, the
modified SW algorithm is not efficient.  The problem is that large clusters of spins usually have a
large net field acting on them and are  prevented from flipping by these fields.  An algorithm for
Ising systems with fields that avoids this problem was introduced by Redner, Machta, and
Chayes\cite{ReMaCh,ChMaRe98b}. In this {\em two-replica} cluster algorithm large clusters are
constructed from two replicas of the same system and have no net field acting on them so that they
may be freely flipped. The two-replica cluster algorithm has been applied to study the phase
transition of benzene adsorbed in zeolites~\cite{DuMaSaAu} and is more efficient than the
conventional Metropolis algorithm for locating and simulating the critical point and the phase
coexistence line.  Combined with the replica exchange method of Swendsen and
Wang~\cite{SwWa86}, the two-replica method has been applied to the random field Ising
model~\cite{MaNeCh00}.  The two-replica method is closely related to the geometric cluster Monte Carlo
 method~\cite{DrKr,HeBl96,HeBl98}.

In this paper, we report on a detailed investigation of the
dynamics of the two-replica cluster (TRC) algorithm as applied to the two-dimensional Ising ferromagnetic in a
staggered field  (equivalently, the Ising antiferromagnet in a uniform field). The \trc algorithm
introduced in Ref.
\onlinecite{ReMaCh} has two components that are not required for detailed balance and ergodicity. 
We studied the contribution to the performance of the algorithm of these optional components.   
We find that the complete
\trc algorithm has a very small dynamic exponent $z < 0.4$.   However, we also find that this
small value of
$z$ requires one of the optional components and that this component depends on a special symmetry of Ising
model in a staggered field. This observation leads to the question of whether cluster methods exist for
efficiently simulating more general Ising models with fields.  We investigated other optional components for
the algorithm but these do not lead to acceleration when fields are present.

This paper is organized as follows. In Sec.\ \ref{sec:ma} we introduce the Ising
model in a staggered field and describe the \trc algorithm.  In Sec.\ \ref{sec:methods} we define the
quantities to be measured and how errors are computed.  In Sec.\ \ref{sec:results}  we present the results.
The paper closes in Sec.\ \ref{sec:disc} with a discussion.
 
\section{The Model and Two-Replica Algorithm}
\label{sec:ma}

\subsection{Ising Model in a Staggered Field}

The Hamiltonian for the Ising model in a staggered field is 
\begin{equation}
\label{eq:H}
 \beta\mathcal{H}[\sigma] = - K\sum_{<i, j>}\sigma_{i}\sigma_{j}
-\sum_{i}H_{i}\sigma_{i} 
\end{equation}
where the spin variables, $\sigma_{i}$ take the values $\pm 1$. $K$ is the coupling strength 
and $H_{i}$ is the magnetic field at site $i$. The summation in the first term of Eq. (\ref{eq:H})
is over nearest neighbors on an $L\times L$ square lattice with periodic boundary conditions and
$L$ even. The second summation is over the sites of the lattice. The staggered field is obtained by
setting
$H_{i}=H$ if $i$ is in the even sublattice and $H_{i}=-H$ if $i$ is in the odd sublattice. The
staggered field breaks the up-down symmetry($\sigma_{i} \leftarrow -\sigma_{i}$) of the zero
field Ising model, however two
symmetries remain.  The Hamiltonian is invariant under even translations:
\begin{equation}
\label{eq:even}
 \sigma_{i+r_0} \leftarrow \sigma_{i} \mbox{ for all } i
\end{equation}
with 
$r_0$ any vector in the even sublattice.  The Hamiltonian is also invariant under odd translations together
with a global flip:
\begin{equation}
\label{eq:odd}
\sigma_{i+r_1} \leftarrow -\sigma_{i} \mbox{ for all } i
\end{equation} with 
$r_1$ any vector in the odd sublattice. 

\begin{figure}
\includegraphics{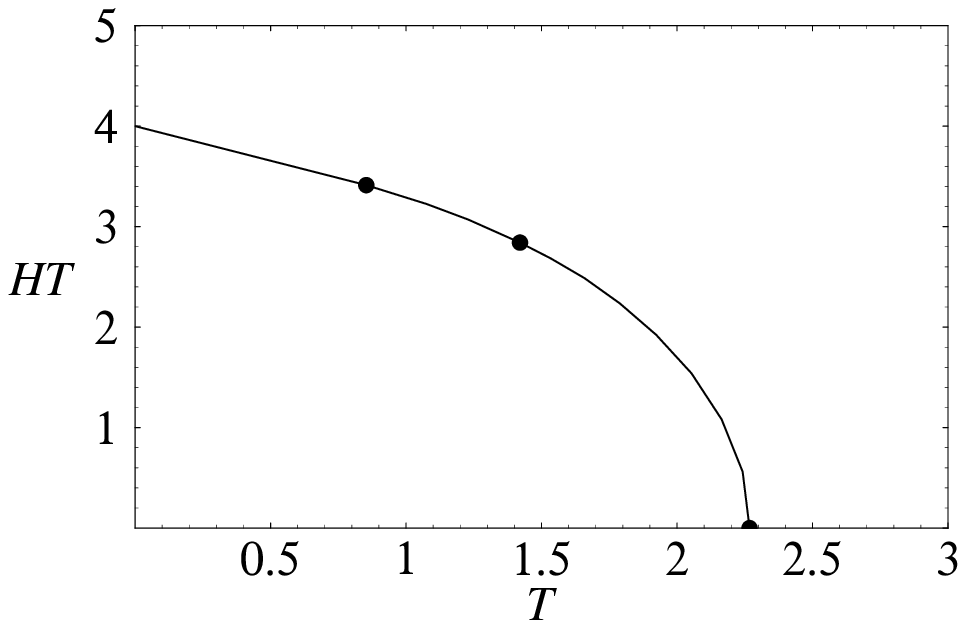}
\caption{Phase diagram of the two-dimensional staggered field Ising model, the three points on the
critical line corresponds respectively to $H$=0, 2, 4.}
\label{fig:phase}
\end{figure}

Figure~\ref{fig:phase} shows the line of critical points, $K_c(H)$ for this model. We carried out simulations at three
points on the critical line taken from the high precision results of Ref.~\onlinecite{BlWu},

\[ K_{c}(0)=0.4406867952\] 
\[ K_{c}(2)=0.7039642053\] 
\[ K_{c}(4)=1.1717153065\]

The basic idea of the two-replica cluster algorithm is to simultaneously simulate two
independent Ising systems, \(\sigma\) and \(\tau\), on the same lattice and in the same field.  Clusters of
pairs of spins in this two-replica system are identified and flipped.  In order to
construct clusters, auxilliary bond variables are introduced. The bond variables
\{$\eta_{ij}$\} are defined for each bond $<i, j>$ and take values 0 and 1. We say that 
$<i,j>$ is {\em occupied} if
$\eta_{ij} = 1$.  A bond $<i,j>$ is {\em satisfied} if $\sigma_i=\sigma_j$ and $\tau_i=\tau_j$.  Only satisfied bonds
may be occupied.  

The two-replica algorithm simulates a joint distribution of the Edwards-Sokal
\cite{EdSo} type for
\{$\sigma_i$\} and
\{$\tau_i$\}, and \{$\eta_{ij}$\}.  The statistical weight
$X[\sigma, \tau, \eta]$ for the joint distribution is
\begin{equation}
\label{eq:X}
 X[\sigma, \tau, \eta]=e^{-G[\sigma,\tau]} \Delta[\sigma, \tau, \eta] B_p[\eta] 
\end{equation}
where 
\begin{equation}
\label{eq:G}
            G=K\sum_{<i, j>} \sigma_i \tau_i \sigma_j \tau_j
                       -\sum_i H_i (\sigma_i + \tau_i) ,
\end{equation}
$B$ is the standard Bernoulli factor,
\begin{equation}
\label{eq:B}
 B_p[\eta] = p^{|\eta|} (1-p)^{N_b-|\eta|} 
\end{equation}
$|\eta|$ = \# $\{<i, j>|\eta_{ij}=1\}$ is the number of occupied bonds and $N_b$ is
the total number of bonds of the lattice. The $\Delta$ factor enforces the
rule that only satisfied bonds are occupied: if for every bond $<i, j>$ such that
$\eta_{ij}=1$ the spins agree in both replicas ($\sigma_{ij}=\sigma_{ij}$
and $\tau_{i}=\tau_{j}$)
then $\Delta[\sigma, \tau, \eta]=1$;
otherwise $\Delta[\sigma, \tau, \eta]=0$.  It is straightforward to show that integrating
$X[\sigma, \tau, \eta]$ over the bond variables, $\eta$ yields the statistical weight for two
independent
Ising model in the same field,
\begin{equation}
\label{eq:weight}
 e^{-\beta\mathcal{H}[\sigma]-\beta\mathcal{H}[\tau]} =
const\sum_{\{\eta\}}X[\sigma, \tau, \eta] 
\end{equation}
if the identification is made that $p=1-e^{-4K}$.

\subsection{Two-Replica Cluster Algorithms}

The idea of the two-replica cluster algorithm is to carry out moves on the spin and bond variables
that satisfy detailed balance and are ergodic with respect to the joint distribution of Eq.\
(\ref{eq:X}).   The occupied bonds $\eta$ define
connected clusters of sites.  We call site
$i$ an \emph{active site} if $\sigma_{i} \neq \tau_{i}$ and clusters are composed either entirely of
active or inactive sites.  If a cluster of active sites  is flipped so that $\sigma \leftarrow -\sigma$
and
$\tau \leftarrow -\tau$ the factor $G$ is unchanged.  

A single Monte Carlo sweep of the TRC algorithm is composed of the following three steps:

\begin{enumerate}
\item Occupy satisfied bond connecting active sites with probability
      $p=1-e^{-4K}$. Identify clusters of active sites connected by occupied bond   
      (including single active sites).  For each cluster $k$, randomly and independently assign a spin value
$s_{k}=\pm1$.
      If site $i$ is in cluster
      $k$ then the new spin values are $\sigma_{i}\leftarrow s_{k}$ and
      $\tau_{i}\leftarrow-s_{k}$. In this way all active sites are updated.

\item Update each replica separately with one sweep of the Metropolis algorithm.

\item Translate the $\tau$ replica by a random amount relative to the
      $\sigma$ replica. If the translation is by an odd vector, all $\tau$
      spins are flipped. 
\end{enumerate}

Step 1 of the \trc is similar to a sweep of the SW algorithm except that clusters are grown in a two-replica system
rather than in a single replica and only active clusters are flipped.  Note also that the bond occupation probability
is  $p=1-e^{-4K}$ for the \trc algorithm and $p=1-e^{-2K}$ for the SW algorithm. It is
straightforward to show that Step 1 of the \trc algorithm satisfies detailed balance with
respect to the joint distribution Eq.\ (\ref{eq:X}). Since only active sites participate in Step 1 of the
algorithm, the Metropolis sweep, Step 2, is required for ergodicity.  Step 3 contains the optional
components of the algorithm: an even translation or an odd translation plus flip of one replica
relative to the other. These moves are justified by the symmetries  of the Ising model in a
staggered field stated in Eqs.\ (\ref{eq:even}) and (\ref{eq:odd}).   When we refer to the
\trc algorithm without further specification, we mean the algorithm described by the Steps 1-3
above.  In the foregoing we also study the \trc with only even translations or with only odd
translations. 

In the \trc algorithm we flip only active clusters but it is also possible to flip inactive clusters if a weight factor
associated with the change in $G$ is used.  We call a flip of an active cluster to
an active cluster ($+-$ to $-+$ or
$-+$ to
$+-$)  an \emph{active flip}. The
\trc algorithm {\em with inactive flips} is obtained by replacing Step 1 with the following: 

 \begin{enumerate}
      \item[1$^\prime$.] Occupy  satisfied bonds with probability $p=1-e^{-4K}$.  Identify clusters connected by
occupied bonds (including single sites).  For each cluster $k$, taken one at a time, randomly propose two new
spin values values, $s_{k}=\pm1$ and $t_{k}=\pm1$ for the $\sigma$ and $\tau$ spins respectively.  Compute
$\delta G$, the change in $G$ that would occur if the spins in the $k^{th}$ cluster are changed to the proposed
values leaving spins in other clusters fixed.  If $\delta G \leq 0$ accept the proposed spin values (set
$\sigma_i\leftarrow s_k$ and $\tau_i\leftarrow t_k$ for all sites $i$ in cluster $k$), otherwise, if
$\delta G > 0$ accept the proposed spin values with probability $e^{-\delta G}$.  
\end{enumerate}

Step 1$^\prime$ is by itself ergodic however it may be useful to add Metropolis sweeps and translations.

\section{Methods}
\label{sec:methods}

We measured three observables using the \trc algorithm: the
absolute value of the magnetization of a single replica, \emph{m}; the energy of a single
replica, $\mathcal{E}$;  and the absolute value of the net staggered magnetization for both
replicas, \emph{s}, 
where the definition of \emph{s} is
\begin{equation}
\label{eq:S}
    \emph{s}=|(\sum_{i \in odd} - \sum_{i \in even})(\sigma_{i}+\tau_{i}) | .
\end{equation}
Note that the
staggered magnetization is conserved by all components of the \trc algorithm except 
Metropolis sweeps and inactive flips. 
For each of these observables we computed expectation values of the integrated
autocorrelation time, $\tau_{int}$ and the exponential autocorrelation time,
$\tau_{exp}$. From $\tau_{int}$, we estimated the dynamic exponent $z$. 

The autocorrelation function for $\phi$, $\Gamma_{\phi\phi}(t)$ is given by,
\begin{equation}
\label{eq:auto-func}
    \Gamma_{\phi\phi}(t)=\lim_{l \rightarrow \infty}\frac{\sum_{t^\prime=1}^{l-t}
(\phi(t^\prime)-\hat{\phi})(\phi(t^\prime+t)-\hat{\phi})}
{\sum_{t^\prime=1}^{l}(\phi(t^\prime)-\hat{\phi})^{2}}.
\end{equation}
The integrated autocorrelation time for observable $\phi$ is defined by
\begin{equation}
\label{eq:in-auto1}
     \tau = \frac{1}{2} + \lim_{t^* \rightarrow \infty}\sum_{t=1}^{t^*}\Gamma_{\phi\phi}(t) 
\end{equation}
and the exponential autocorrelation time for an observable $\phi$ is defined by
\cite{SaSo97}
\begin{equation}
\label{eq:ex-auto}
   \tau_{exp, \phi}=\lim_{t \rightarrow \infty} \frac{-|t|}{\log
   \Gamma_{\phi \phi}(t)} . 
\end{equation}

In practice the limits in Eqs.\  (\ref{eq:auto-func}), (\ref{eq:in-auto1}) and (\ref{eq:ex-auto})  must be
evaluated at finite values.  The length of the Monte Carlo runs determine $l$ and are discussed below.
Following   Ref.~\onlinecite{SaSo97}, we define
\begin{equation}
\label{eq:in-auto2}
     \tau_{int, \phi}=\frac{1}{2} + \sum_{t=1}^{t^{*}}\Gamma_{\phi\phi}(t) 
\end{equation}
and choose the cutoff $t^{*}$ to be the smallest integer such that  
$t^{*} \geq \kappa \tau_{int, \phi}$, where $\kappa$ = 6.  We used the least-squares method to fit  $\log
\Gamma_{\phi\phi}(t/\tau_{int, \phi})
$ as a function of $t$ to obtain the ratio of $\tau_{int, \phi}/\tau_{exp, \phi}$ and chose a cut-off at
$t/\tau_{int,\phi}=5$.

We used the blocking method~\cite{NeBa99,SaSo97} to estimate errors. The whole
sample of $n$ MC measurements was divided into $m$  blocks of equal length $l=n/m$. For each
block $i$ and each measured quantity $A$, we computed the mean  $\hat{A}_i$ . Our estimates of 
$\hat{A}$ and its error $\delta A$ are obtained from:
\begin{equation}
\label{eq:average}
    \hat{A}=\frac{1}{m} \sum_{i=1}^{m}\hat{A}_{i}  
\end{equation}
\begin{equation}
\label{eq:error}
    \delta\hat{A}^2=\frac{1}{m(m-1)}\sum_{i=1}^{m}(\hat{A
}-\hat{A}_{i})^{2}  
\end{equation}
In our simulations, we divided the whole sample into $m$ blocks where $m$ is between 10 and 30. 

For the data
collected using the \trc algorithm, each block has a length
$l \geq 10^{3}\tau_{int}$. For the data collected using modifications of the \trc algorithm, each block has a
length $l \geq 10^{2}\tau_{int}$.  Data were collected for $H=0$, 2 and 4 and for size $L$ in the range 16 to
256.

\section{Results}
\label{sec:results}

\subsection{Integrated Autocorrelation Time}

Table~\ref{tab:m&s&e} gives the integrated autocorrelation time using the \trc algorithm
for the magnetization, energy and staggered magnetization. 
Table~\ref{tab:m&s&e} is comparable to the Table in Ref.\ \onlinecite{ReMaCh} but
the present numbers are systematically larger, especially at the larger system sizes. 
This discrepancy may be due to the sliding cut-off $t^{*}$ used here instead of a fixed cut-off at 200
employed in Ref.\  \onlinecite{ReMaCh}.  

Table~\ref{tab:odd&even}  gives the integrated autocorrelation times
for magnetization using the
\trc with  only even or only odd translations.  The comparison of \trc algorithm with only even translations
and with only odd translations in Tables~\ref{tab:odd&even} shows that odd translations together with
global flips of one replica relative to another are essential to achieve small and slowly growing
autocorrelation times when the staggered field is present.  

Table~\ref{tab:compare} shows the magnetization autocorrelation times using different algorithms for system
size $L=80$. The Swendsen-Wang (SW) algorithm has the smallest $\tau_{int,m}$ in the absence of fields.
However, when fields are present and the SW algorithm is then modified  according to the method of
Ref.~\onlinecite{DoSeTa} the performance is worse even than that of the Metropolis algorithm.  The slow
equilibration of the SW algorithm in the presence of the staggered field is due to small acceptance
probabilities for flipping large clusters.  On the other hand, the presence of staggered fields does not
significantly change the performance the two-replica algorithm so long as odd translations are present.  
Inactive flips are helpful when there is no staggered field but when the staggered field is turned on, the
autocorrelation time is not substantially improved by inactive flips. The explanation for the ineffectiveness of
inactive flips when the staggered field is present is that the probability of accepting an inactive flip is small.
For example, this probability is
$1.4\%$ for $L=80$ and $H=4$.

The CPU time per spin on a Pentium III 450 MHz
machine was also measured for the various algorithms and is listed in Table~\ref{tab:compare}
for $L=80$ .  By considering a range of system sizes we found that  the CPU time for one MC sweep
of the \trc algorithm  increases nearly linearly with the number of spins.   The \trc algorithm is a factor of
3 slower than the Metropolis algorithm but this difference is more than compensated for by system size
$80$ by the much faster equilibration of the \trc algorithm.  Even without odd translations, the \trc
algorithm outperforms Metropolis for size 80. 

\subsection{Exponential Autocorrelation Time}

The ratio of the integrated to exponential autocorrelation times was found to be nearly independent of
system size over the range $L=16$ to $L=256$. We found that over this size range
$\tau_{int, m}/\tau_{exp, m}$ varied from
$0.448\pm0.008$ to $0.425\pm0.008$ for $H=0$; from $0.44\pm0.01$ to $0.43\pm0.01$
for $H=2$; and from $0.448\pm0.009$ to $0.409\pm0.009$ for $H=4$.  The ratio
$\tau_{int, s}/\tau_{exp, s}$ is also nearly independent of $L$ and $H$ and is about 0.45. 
The ratio
$\tau_{int,\e}/\tau_{exp,
\e}$ is nearly  independent of $L$ but decreases slowly with $H$ ranging from 0.29 to 0.25
as $H$ ranges from 0 to 4.  The almost constant $\tau_{int,\phi}/\tau_{exp, \phi}$ for different sizes suggests
that the integrated and exponential autocorrelation times are governed by the same dynamic critical
exponent.

\subsection{Dynamic Exponent}
\begin{figure}
\includegraphics{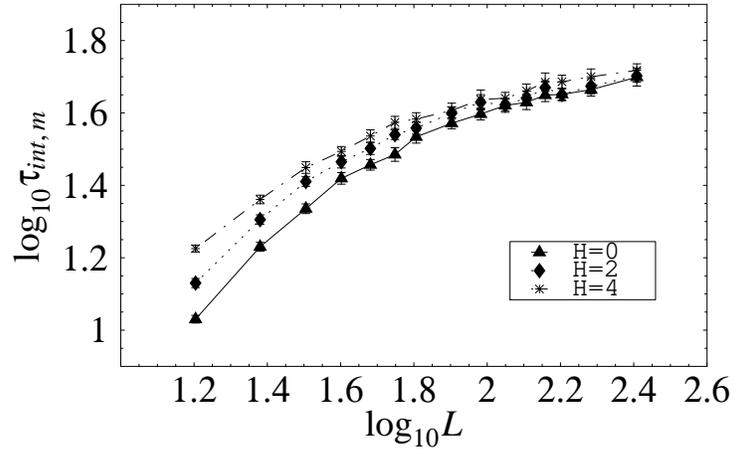}
\caption{Logarithm of the magnetization autocorrelation time $\tau_{int,m}$  vs.\ logarithm of system size for
 $H=0$, 2, 4.}
\label{fig:m-bilog}
\end{figure}

\begin{figure}
\includegraphics{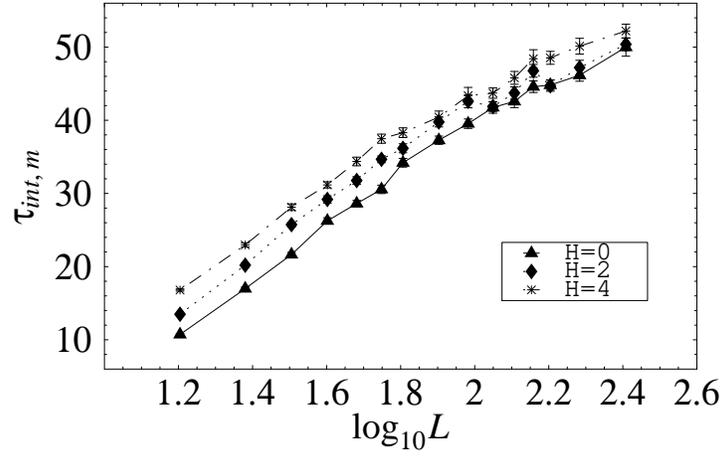}
\caption{Magnetization autocorrelation time $\tau_{int,m}$ vs.\ logarithm of system size $L$ for
 $H=0$, 2, 4.}
\label{fig:m-log}
\end{figure}

\begin{figure}
\includegraphics{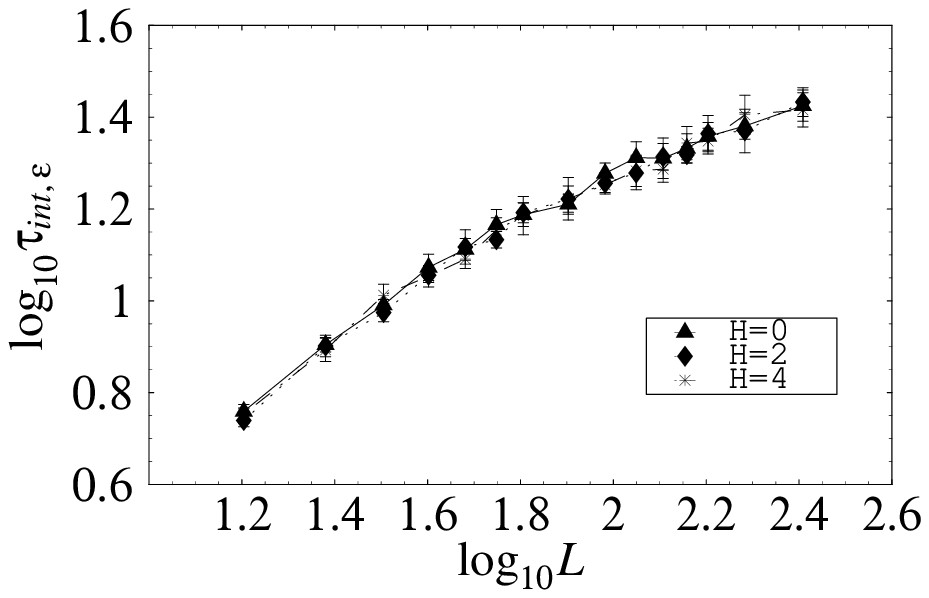}
\caption{Logarithm of energy autocorrelation time $\tau_{int,\e}$ vs.\ logarithm of system size $L$
for $H=0$, 2, 4.}
\label{fig:e-bilog}
\end{figure}

\begin{figure}
\includegraphics{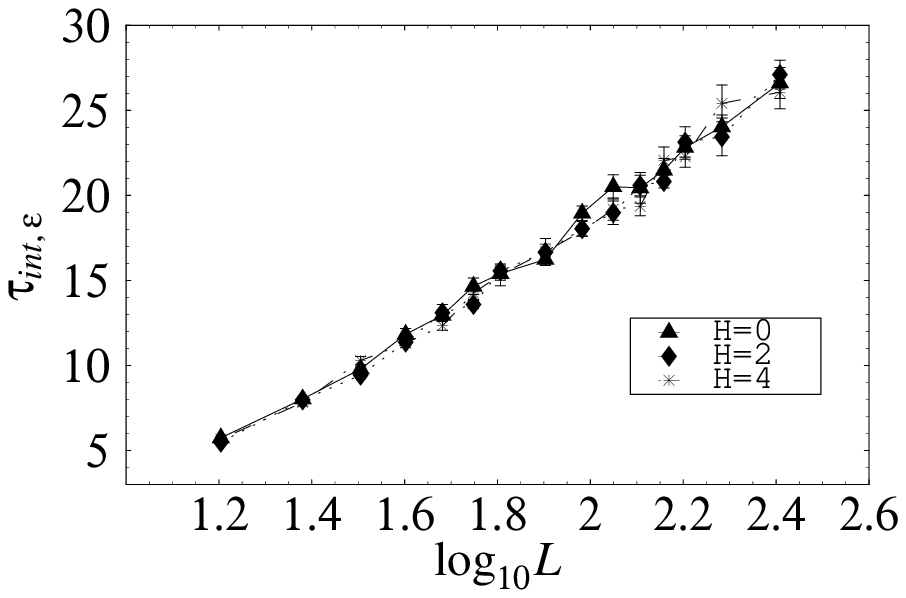}
\caption{ Energy autocorrelation time $\tau_{int,\e}$ vs.\ logarithm of system
size $L$ for $H=0$, 2, 4.}
\label{fig:e-log}
\end{figure}

\begin{figure}
\includegraphics{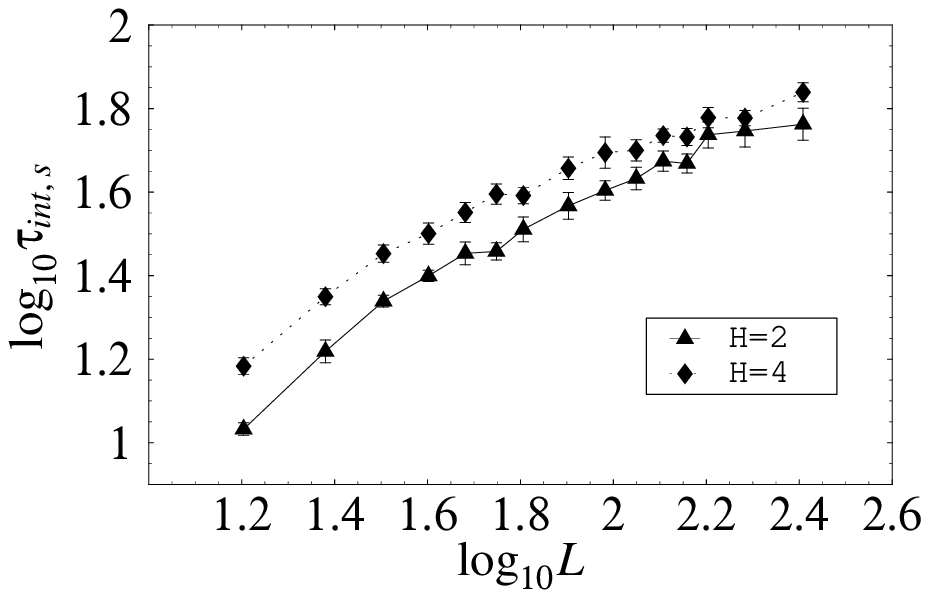}
\caption{Logarithm of staggered magnetization autocorrelation time $\tau_{int,s}$ vs.\  logarithm of
system size
$L$ for $H=0$, 2, 4.}
\label{fig:s-bilog}
\end{figure}

\begin{figure}
\includegraphics{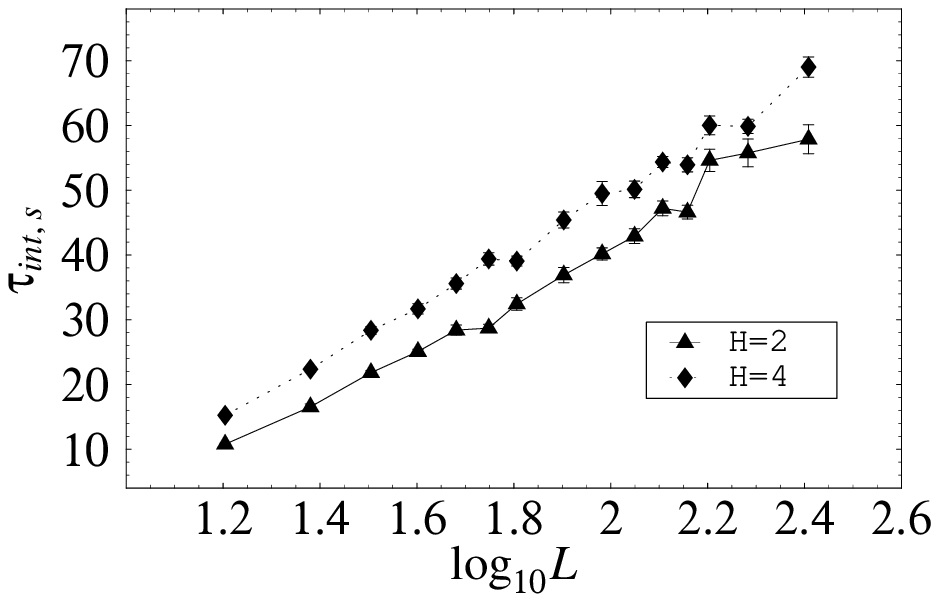}
\caption{Staggered magnetization autocorrelation time $\tau_{int,s}$ vs.\ logarithm of system
size $L$ for $H=0$, 2, 4.}
\label{fig:s-log}
\end{figure}

Figures~\ref{fig:m-bilog} and \ref{fig:m-log}  show the magnetization integrated autocorrelation
time for the \trc plotted on log-log and log-linear scales, respectively.  Figures~\ref{fig:e-bilog} and
\ref{fig:e-log} show the energy integrated autocorrelation time for the
\trc plotted on log-log and log-linear scales, respectively. Figures~\ref{fig:s-bilog} and \ref{fig:s-log} show the
staggered magnetization integrated autocorrelation time for the
\trc plotted on log-log and log-linear scales, respectively.

\begin{figure}
\includegraphics{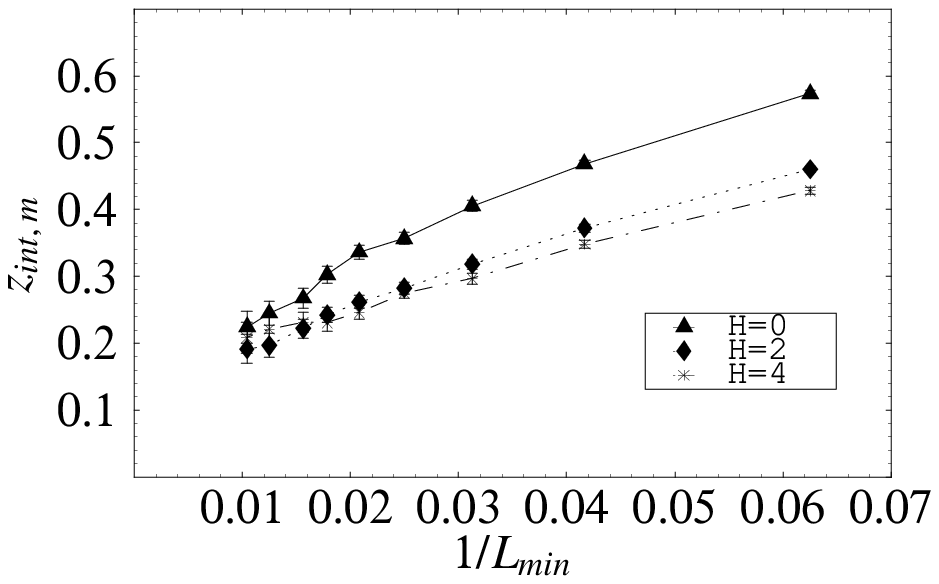}
\caption{Finite size dynamic critical exponent for magnetization $z_{int,m}$ vs.\ the reciprocal of the
minimum size $L_{min}$ used in the fit. }
\label{fig:m-z}
\end{figure}

\begin{figure}
\includegraphics{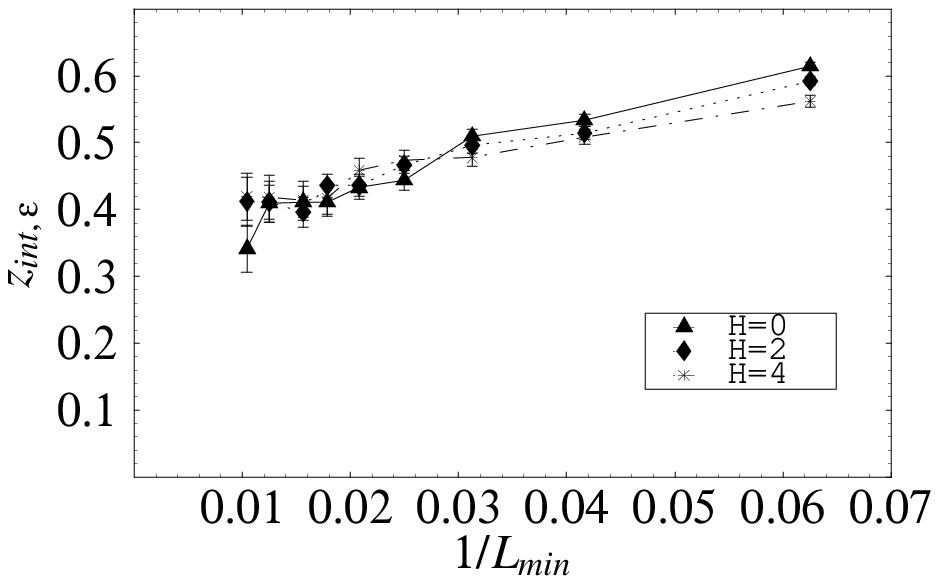}
\caption{Finite size dynamic critical exponent for energy $z_{int,\e}$ vs.\ the reciprocal of the
minimum size $L_{min}$ used in the fit.
translations.}
\label{fig:e-z}
\end{figure}

\begin{figure}
\includegraphics{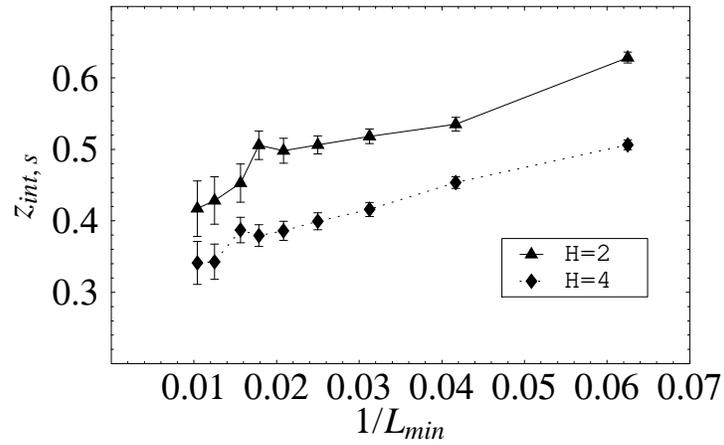}
\caption{Finite size dynamic critical exponent for staggered magnetization $z_{int,s}$ vs.\ the reciprocal of the
minimum size $L_{min}$ used in the fit.
translations.}
\label{fig:s-z}
\end{figure}

\begin{figure}
\includegraphics{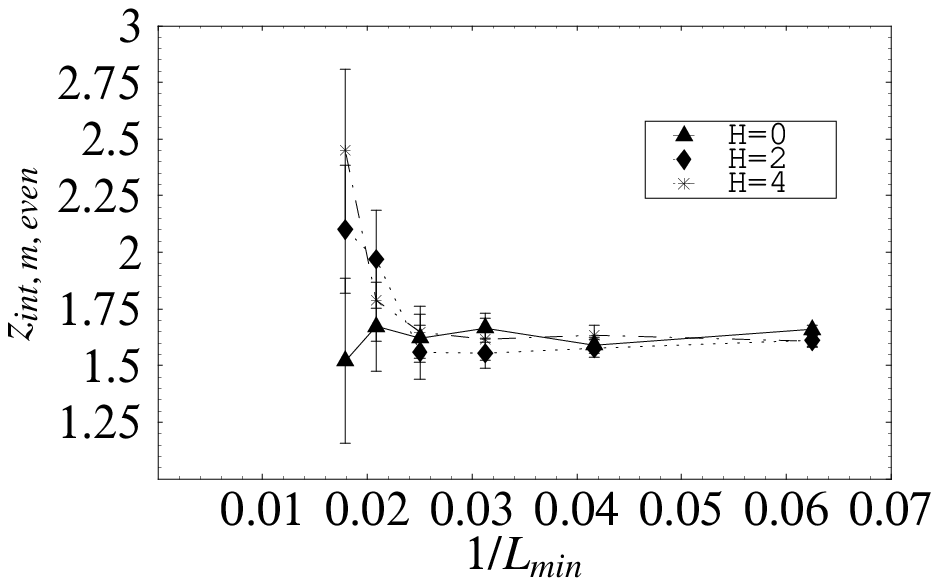}
\caption{Finite size dynamic critical exponent for magnetization $z_{int,m,even}$ vs.\ the reciprocal of the
minimum size $L_{min}$ used in the fit for the \trc with only even
translations.}
\label{fig:even-z}
\end{figure}

\begin{figure}
\includegraphics{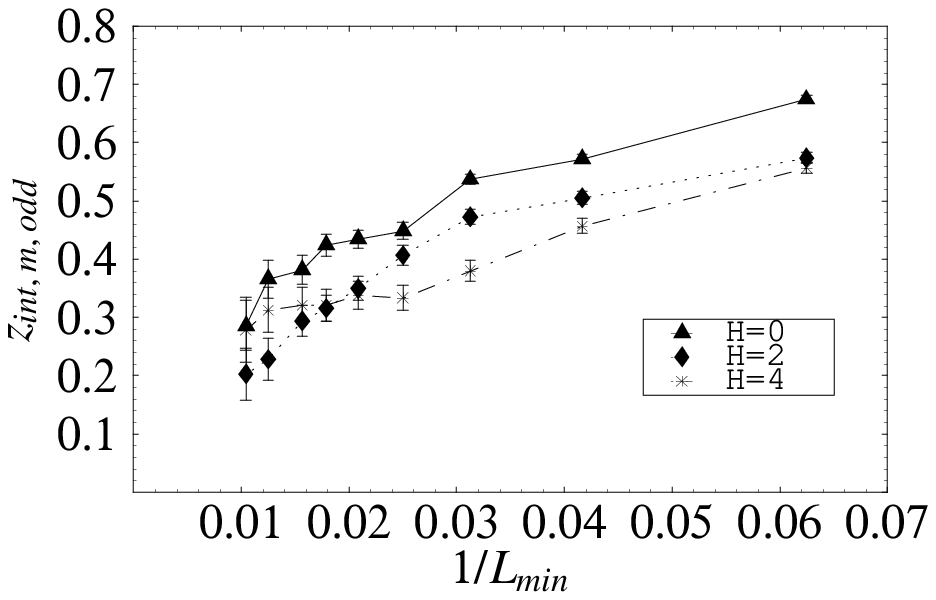}
\caption{Finite size dynamic critical exponent for magnetization $z_{int,m,odd}$ vs.\ the reciprocal of the
minimum size $L_{min}$ used in the fit for the \trc with only odd
translations.}
\label{fig:odd-z}
\end{figure}

For the
whole range of $L$, logarithmic growth appears to give a somewhat better fit than a simple power law,
particularly for the magnetization.  Therefore, our results are consistent with $z=0$ for the \trc algorithm. 
Under the assumption that the dynamic exponent is not zero, we also carried out weighted least-squares fits
to the form
$A L^{z}$ and varied $L_{min}$, the minimum system size included in the fit.  
Figures~\ref{fig:m-z},  \ref{fig:e-z} and \ref{fig:s-z} show the dynamic
exponent $z$ for the magnetization, energy and staggered magnetization, respectively, as a function of 
$L_{min}$ using the \trc algorithm.  Figures \ref{fig:even-z} and \ref{fig:odd-z},  show the dynamic
exponent as a function of $L_{min}$ for the magnetization for the \trc with only even translations and
only odd translations, respectively.  In all cases except $z_{int,m,even}$, the dynamic exponent is a
decreasing function of $L_{min}$.  For the magnetization, $z_{int,m}$ appears to extrapolate to a value
between 0.1 and 0.2 as $L_{min} \rightarrow \infty$ while for the energy and staggered magnetization, the
dynamic exponent appears to extrapolate to a value between 0.3 and 0.4.  The small value of the
dynamic exponent requires that odd translations and flips are included in the algorithm.  From
Fig.\ \ref{fig:even-z} it is clear that the dynamic exponent is near 2 for the
\trc algorithm with only even translations. 

Table \ref{tab:dyn-exp} gives results of the weighted least squares fits for $z$ for the smallest
values of $L_{min}$ for which there is a reasonable confidence level.   Since there is a general
downward curvature in the log-log graphs, these numbers are likely to be overestimates of the
asymptotic values.  Thus, we can conclude that the asymptotic dynamic exponent for the \trc
algorithm is  likely to be less than $0.4$ and is perhaps exactly zero.  The dynamic exponent is
apparently independent of the strength of the staggered field.   For the case of the SW algorithm
applied to the two-dimensional Ising with no staggered field the best estimate is
$z=0.25\pm0.01$\cite{BaCo,CoBa} but the results are also consistent with logarithmic growth of
relaxation times.  The numbers for dynamic exponent for the SW appear to be smaller than for
the \trc algorithm but this difference may simply reflect larger corrections to scaling in the case
of the \trc. 

\section{Discussion}
\label{sec:disc}
We studied the dynamics of the two-replica cluster algorithm applied to the two-dimensional
Ising model in a staggered field. We found that the dynamic exponent of the algorithm is either
very small ($z \leq 0.4$) or zero ( $\tau \sim \log L$) and that the dynamic exponent does not
depend on the strength of the staggered field.  A precise value of $z$ could not be determined
because of large corrections to scaling.  We tested the importance of various optional components
of the algorithm and found that an odd translation and global flip of one replica relative to
another is essential for achieving rapid equilibration.  Without this component,
$z$ is near 2 so there is no qualitative improvement over the Metropolis
algorithm.  An odd translation and global flip of one replica relative to the other allows for a
large change of the total magnetization of the system with an acceptance fraction of
$100\%$.  Large changes in the global magnetization may also occur in the Swendsen-Wang
algorithm in a field or via inactive flips in the \trc algorithm but these flips have a small
acceptance fraction due to the staggered field.  Unfortunately, the odd translation and flip move
is allowed because of a special symmetry of the Ising model in a staggered field.  For more
general Ising systems with translationally invariant fields, we expect performance similar to the
\trc with even translations only.  In this case,  the autocorrelation time is significantly less
than for the Metropolis algorithm but the dynamic exponent is about the same.  While the
two-replica approach is useful for these more general problems of Ising systems with fields, it
does not constitute a method that overcomes critical slowing down except when additional
symmetries are present that allow one replica to be flipped relative to the other. 
Development of general methods for efficiently simulating critical spin systems with fields
remains an open problem.

\acknowledgements
This work was supported in part by NSF grants DMR 9978233.

\newpage




\newpage

\pagebreak

\printtables

\begin{table}   
 \caption{Integrated autocorrelation times for the \trc algorithm for the 
          magnetization of a single replica $\tau_{m}$, the net staggered
          magnetization of both replicas $\tau_{s}$ and the 
          energy of a single replica $\tau_{e}$.}

\vspace{0.2in}

 \begin{tabular}{|c|c|c|c|c|c|c|c|c|} \hline\hline
   \multicolumn{1}{|c|}{} &
   \multicolumn{2}{c|}{ $H$ = 0 } &
   \multicolumn{3}{c|}{ $H$ = 2 } &
   \multicolumn{3}{c|}{ $H$ = 4 } \\  \cline{2-9}
    \raisebox{2.25ex}[0pt]{L(size)}& $\tau_{int, m}$ & $\tau_{int, \e}$ &
      $\tau_{int, m}$ & $\tau_{int, s}$ & $\tau_{int, \e}$ &
      $\tau_{int, m}$ & $\tau_{int, s}$ & $\tau_{int, \e}$
\\ \hline

16 & $10.7\pm0.1$ & $5.73\pm0.09$ & $13.5\pm0.2$ & $10.8\pm0.2$ & $5.49\pm0.08$ &
$16.8\pm0.2$ & $15.2\pm0.3$ & $5.7\pm0.1$ \\
24& $17.0\pm0.2$ & $8.0\pm0.1$ & $20.2\pm0.3$ & $16.5\pm0.5$ & $8.0\pm0.2$ &
$23.0\pm0.3$ & $22.4\pm0.4$ & $7.8\pm0.2$ \\
32& $21.6\pm0.3$ & $9.8\pm0.1$ & $25.7\pm0.3$ & $21.8\pm0.3$ & $9.4\pm0.2$ &
$28.1\pm0.4$ & $28.4\pm0.6$ & $10.3\pm0.2$ \\
40& $26.3\pm0.4$ & $11.8\pm0.3$ & $29.2\pm0.5$ & $25.1\pm0.3$ & $11.4\pm0.2$ &
$31.4\pm0.4$ & $31.7\pm0.8$ & $11.3\pm0.3$ \\
48& $28.6\pm0.4$ & $12.9\pm0.3$ & $31.8\pm0.5$ & $28.4\pm0.8$ & $13.1\pm0.5$ &
$34.4\pm0.6$ & $35.6\pm0.9$ & $12.3\pm0.3$ \\
56& $30.6\pm0.6$ & $14.7\pm0.5$ & $34.7\pm0.5$ & $28.7\pm0.6$ & $13.6\pm0.2$ &
$37.5\pm0.6$ & $39\pm1$ & $14.3\pm0.4$  \\
64& $34.2\pm0.6$ & $15.4\pm0.4$ & $36.2\pm0.6$ & $32\pm1$ & $15.6\pm0.3$ &
$38.1\pm0.5$ & $39.1\pm0.8$ & $15.3\pm0.6$ \\
80& $37.3\pm0.6$ & $16.2\pm0.4$ & $39.8\pm0.7$ & $37\pm1$ & $16.7\pm0.5$ &
$40.4\pm0.8$ & $45\pm1$ & $16.7\pm0.8$  \\
96& $39.5\pm0.6$ & $19.0\pm0.4$ & $42.6\pm0.9$ & $40.2\pm0.9$ & $18.0\pm0.4$ &
$43\pm1$ & $50\pm2$ & $17.9\pm0.3$  \\
112 & $41.7\pm0.8$ & $20.5\pm0.7$ & $41.9\pm0.7$ & $43\pm1$ & $19.0\pm0.7$ &
$43.7\pm0.7$ & $50\pm1$ & $19.2\pm0.7$  \\
128& $42.6\pm0.8$ & $20.4\pm0.9$ & $43.7\pm0.8$ & $47\pm1$ & $20.6\pm0.6$ &
$45.8\pm0.9$ & $54.4\pm0.9$ & $19.4\pm0.6$ \\
144& $44.6\pm0.8$ & $21.5\pm0.7$ & $46.8\pm0.9$ & $47\pm1$ & $20.8\pm0.4$ &
$48\pm1$ & $54\pm1$ & $22.1\pm0.8$  \\
160& $44.8\pm0.7$ & $22.8\pm0.7$ & $44.7\pm0.8$ & $55\pm2$ & $23.1\pm0.9$ & 
$48.5\pm0.9$ & $60\pm1$ & $22.3\pm0.6$ \\   
192& $46.2\pm0.8$ & $24.0\pm0.7$ & $47.2\pm0.9$ & $56\pm2$ & $23\pm1$ &
$50\pm1$ & $60\pm1$ & $25\pm1$ \\
256& $50\pm1$ & $26.6\pm0.9$ & $50.4\pm0.9$ & $58\pm2$ & $27.1\pm0.8$ &
$52.2\pm0.9$ & $69\pm2$&1 $26\pm1.0$  \\

\hline\hline
\end{tabular}
\label{tab:m&s&e}

\end{table}

\begin{table}
 \caption{Integrated autocorrelation times for the 
  magnetization of a single replica using the \trc algorithm with odd translations only, $\tau_{int,
m, odd}$  and even translations only, $\tau_{int, m, even}.$}

\vspace{.2in}

\begin{tabular}{|c|c|c|c|c|c|c|} \hline\hline
   \multicolumn{1}{|c|}{} &
   \multicolumn{2}{c|}{$H$=0} &
   \multicolumn{2}{c|}{$H$=2} &
   \multicolumn{2}{c|}{$H$=4} \\ \cline{2-7}
    \raisebox{2.25ex}[0pt]{L(size)}& 
      $\tau_{int, m, odd}$  &  $\tau_{int, m, even}$ &
      $\tau_{int, m, odd}$  &  $\tau_{int, m, even}$ &
      $\tau_{int, m, odd}$  &  $\tau_{int, m, even}$

\\ \hline

16 & $ 11.6\pm0.2 $ & $ 12.6\pm0.1 $ & $ 14.8\pm0.4 $ & $ 17.8\pm0.3 $ & $ 19.0\pm0.3 $
& $ 24.6\pm0.5 $  \\
24 & $ 19.0\pm0.4 $ & $ 25.7\pm0.3 $ & $ 24.1\pm0.5 $ & $ 34.9\pm0.8 $ & $ 28.4\pm0.5 $
& $ 46\pm1 $\\
32 & $ 25.9\pm0.3 $ & $ 39.3\pm0.8 $ & $ 30.3\pm0.4 $ &  $ 56\pm2 $ & $ 35\pm1 $ & $
77\pm5 $ \\
40 & $ 32\pm1 $ & $ 58\pm2 $ &  $ 34.5\pm0.8 $ & $ 83\pm4 $ & $ 44\pm1 $ & $  109\pm4 $ 
\\
48 & $ 36.6\pm1.0 $ & $ 77\pm4 $ & $ 39\pm1 $ & $ 100\pm7$ & $ 44\pm2 $& $  144\pm6 $ 
\\
56 & $ 39.1\pm0.8 $ & $ 106\pm9 $ & $ 45\pm1 $ & $ 128\pm6 $ & $ 49\pm1 $ & $188\pm17
$ \\
64 & $ 44\pm1 $ & $ 126\pm8 $ & $ 48\pm1 $ & $ 166\pm11 $ & $ 51\pm2 $ & $ 203\pm12 $
\\
80 & $ 46\pm1 $ & $ 186\pm18 $ & $ 54\pm2 $ & $ 283\pm27 $ & $ 55\pm1 $ & $434\pm43 $ 
\\
96 & $ 55\pm2 $ & $-$ & $ 57\pm1 $ & $-$  & $ 59\pm2 $ & $-$ \\
112 & $ 54\pm2 $ & $-$ & $ 60\pm3 $ & $-$ & $ 63\pm3 $ & $-$ \\
128 & $ 58\pm2 $ & $-$ & $ 62\pm3 $ & $-$ & $ 66\pm3 $ & $-$ \\ 
144 & $ 63\pm2 $ & $-$ & $ 65\pm2 $ & $-$ & $ 68\pm3 $ & $-$ \\
160 & $ 64\pm1 $ & $-$ & $ 68\pm3 $ & $-$ & $ 70\pm3 $ & $-$ \\
196 & $ 68\pm3 $ & $-$ & $ 67\pm2 $ & $-$ & $ 75\pm4 $ & $-$ \\
256 & $ 69\pm2 $ & $-$ & $ 68\pm3 $ & $-$ & $ 77\pm4 $ & $-$ \\

\hline\hline
\end{tabular}
\label{tab:odd&even}
\end{table}

\begin{table}[p]
\caption{Magnetization integrated autocorrelation times and CPU times for several 
  algorithms for $L=80$.}

\vspace{.2in} 

\begin{tabular}{|c|c|c|c|c|}\hline\hline
   \multicolumn{1}{|c|}{} &
   \multicolumn{3}{c|}{Integrated Autocorrelation Time} &
   \multicolumn{1}{c|}{ CPU time } \\ \cline{2-4}
     \raisebox{2.25ex}[0pt]{Algorithm} & $H=0$ & $H=2$ & $H=4$ &
($10^{-6}$
sec/sweep/spin)

\\ \hline

TRC & $37.3\pm0.6$ & $39.8\pm0.7$ & $40.4\pm0.8$ & 3.1 \\ \hline
TRC & & & & \\  
odd translations only & \raisebox{2.25ex}[0pt]{$46\pm1$} &
\raisebox{2.25ex}[0pt]{$54\pm2$} & \raisebox{2.25ex}[0pt]{$55\pm1$} &
\raisebox{2.25ex}[0pt]{3.0} \\ \hline
TRC & & & & \\
even translations only &\raisebox{2.25ex}[0pt]{ $186\pm18$} &
\raisebox{2.25ex}[0pt]{$283\pm27$} & \raisebox{2.25ex}[0pt]{$435\pm43$} &
\raisebox{2.25ex}[0pt]{2.9} \\ \hline
TRC \& inactive flips & & & & \\
even translations only & \raisebox{2.25ex}[0pt]{$33.6\pm0.9$} &
\raisebox{2.25ex}[0pt]{$246\pm27$} & \raisebox{2.25ex}[0pt]{$372\pm23$} &
\raisebox{2.25ex}[0pt]{4.6} \\ \hline 
TRC & & & & \\
no translations &\raisebox{2.25ex}[0pt]{$335\pm18$} &
\raisebox{2.25ex}[0pt]{$440\pm24$} &
\raisebox{2.25ex}[0pt]{$773\pm47$} & \raisebox{2.25ex}[0pt]{2.6} \\ \hline
Swendsen-Wang & $4.12\pm0.02$ & $4682\pm173$ & $5707\pm48$ & 1.3 \\ \hline
Metropolis & $928\pm99$ & $1892\pm158$ & $2959\pm236$ & 1.1\\

\hline\hline
\end{tabular}
\label{tab:compare}
\end{table}

\begin{table}[p]
\caption{Estimated dynamic exponents together with minimum 
size used in the fit and confidence level  for  the TRC algorithm, TRC algorithm with odd
translation only and TRC algorithm with even translation only.}

\vspace{.2in}   

\begin{tabular}{|c|c|c|c|}\hline\hline
  \multicolumn{1}{|c|}{dynamic exponent $z$} &
  \multicolumn{1}{c|}{$H=0$} &
  \multicolumn{1}{c|}{$H=2$} &
  \multicolumn{1}{c|}{$H=4$} \\ \hline

$z_{int, m}$ & $0.20\pm0.03$ & $0.20\pm0.02$ & $0.23\pm0.01$ \\
$(L_{min}, level)$ & $(112, 86\%)$ & $(80, 12\%)$ & $(56, 75\%)$ \\ \hline
$z_{int, \e}$ & $0.34\pm0.04$ & $0.40\pm0.02$ & $0.42\pm0.02$ \\
$(L_{min}, level)$ & $(80, 95\%)$ & $(64, 82\%)$ & $(56, 60\%)$ \\ \hline
$z_{int, s}$ &  & $0.42\pm0.03$ & $0.34\pm0.02$ \\
$(L_{min}, level)$ & \raisebox{2.25ex}[0pt]{---} & $(80, 3\%)$ & $(80, 14\%)$ \\ \hline
$z_{int, m, odd}$ & $0.17\pm0.07$ & $0.23\pm0.04$ & $0.33\pm0.02$ \\
$(L_{min}, level)$ & $(144, 89\%)$ & $(80, 60\%)$ & $(40, 95\%)$ \\ \hline
$z_{int, m, even}$ & $1.67\pm0.06$ & $1.97\pm0.22$ & \\
$(L_{min}, level)$ & $(32, 96\%)$ & $(48, 71\%)$ & \raisebox{2.25ex}[0pt]{---}\\
\hline\hline   
\end{tabular}
\label{tab:dyn-exp}
\end{table}

\end{document}